\documentclass[aps,prl,floatfix,floats,twocolumn,letterpaper,groupedaddress,
showpacs,twoside]{revtex4}
\usepackage{graphicx}
\usepackage{color}
\usepackage{ulem}
\newcommand{\la}{\langle}
\newcommand{\ra}{\rangle}
\newcommand{\nicefrac}[2]{\leavevmode\kern.1em
    \raise.5ex\hbox{\the\scriptfont0 #1}\kern-.1em
    /\kern-.15em\lower.25ex\hbox{\the\scriptfont0 #2}}
\begin{document}



\title{Relaxation of Terrace-width Distributions: Physical Information from Fokker-Planck Time}


\author{Ajmi BH. Hamouda$^{a,b}$}
\email[]{hammouda@umd.edu}
\author{Alberto Pimpinelli$^{a,b}$}
\email[]{alpimpin@univ-bpclermont.fr}
\author{T. L. Einstein$^b$}
\email[]{einstein@umd.edu}
\affiliation{$^{\textcolor{black}{a}}$Department of Physics, University of Maryland, College Park, Maryland 20742-4111 USA
\\ $^{\textcolor{black}{b}}$LASMEA, UMR 6602 CNRS/Universit\'e Blaise Pascal -- Clermont 2,
F-63177 Aubi\`ere cedex, France}

\date{\today}

\begin{abstract}
 Recently some of us have constructed a Fokker-Planck formalism to describe the equilibration of the terrace-width distribution of a vicinal surface from an arbitrary initial configuration.  However, the meaning of the associated relaxation time, related to the strength of the random noise in the underlying Langevin equation, was rather unclear.  Here we present a set of careful kinetic Monte Carlo simulations that demonstrate convincingly that the time constant shows activated behavior with a barrier that has a physically plausible dependence on the energies of the governing microscopic model. Furthermore, the Fokker-Planck time at least semiquantitatively tracks the actual physical time.
\end{abstract}


\maketitle
\section{Introduction}
With equilibrium properties of vicinal surfaces---especially the form of the terrace width distribution (TWD)---now relatively well understood \cite{tle},
much attention is focusing on non-equilibrium aspects, which have long been of interest.
In a previous paper some of us \cite{pge} derived the following
Fokker-Planck (FP) equation [Eq.~(\ref{eq:fp3})] to describe the distribution of spacings between steps on a vicinal surface during relaxation to equilibrium.  The goal was to describe the relaxational evolution of this spacing distribution rather than the evolution of the positions of individual steps as in a previous investigation \cite{RV,ncb,mu,jdw}.  As in all those papers, we simplify to a one-dimensional (1D) model, in which a step is represented by its position in the $\hat{x}$ direction (the downstairs direction in ``Maryland" notation), averaged over the $\hat{y}$ direction (along the mean direction of the step, the ``time-like" direction in fermionic formulations) \cite{hpe1}.  This picture implicitly assumes that one is investigating time scales longer than that of fluctuations along the step.

We started with the Dyson Coulomb gas/Brownian motion model;\cite{D62,NS93} made the mean-field-like assumption, when computing interactions, that all but adjacent steps are
separated by the appropriate integer multiple of the mean spacing; and set the width of the confining [parabolic] potential in the model to
produce a self-consistent solution.  Details are provided in the Appendix, which expands the earlier derivation and corrects some inconsequential errors in intermediate stages \cite{pge}.  We found the following:
\begin{equation}
\label{eq:fp3}
{\partial P(s,\tilde t)\over\partial \tilde t}={\partial \over\partial s}\left[\left({2b_\varrho}s-{\varrho\over s}\right)P(s,\tilde t)
\right]+{}{\partial^2 \over\partial s^2}[P(s,\tilde t)],
\end{equation}
where $s$ is the distance $w$ between adjacent steps divided by its average value $\la w\ra$, determined by the slope of the vicinal surface.

The steady-state solution of Eq.~(\ref{eq:fp3}) has the form of the generalized Wigner surmise (GWS), thus  
\begin{eqnarray}
P_\varrho(s)&=& a_\varrho s^\varrho {\rm e}^{-b_\varrho s^2}
\\ a_\varrho &=&
\frac{2\left[\Gamma\! \left(\! {\varrho \! +\! 2\over2}\right)\right]\! ^{\varrho \! +\! 1}}{
\left[\Gamma\! \left(\! {\varrho \! +\! 1\over2}\right)\right]^{\varrho \! +\! 2}}\qquad
b_\varrho =
\left[\frac{\Gamma\! \left({\varrho \! +\! 2\over2}\right)}
{\Gamma\! \left({\varrho \! +\! 1\over2}\right)}\right]^{2} \nonumber
\label{eq:Ps}
\end{eqnarray}
where the constants $b_\varrho$ and $a_\varrho$
assure unit mean and normalization, respectively.  
(The Wigner surmise, Eq.~(2) 
pertains to the special cases $\varrho$=1,2,4; the generalization is to use this expression for arbitrary $\varrho \ge 1$.)
The dimensionless variable $\varrho$ gauges the strength $A$ of the $A/w^2$ energetic repulsion between steps: $(\varrho - 1)^2 =
1 + 4A\tilde{\beta}/(k_BT)^2$, where $\tilde{\beta}$ is the step stiffness.  The dimensionless FP time $\tilde{t}$ can be
written as $t/\tau$; here the relaxation time $\tau$ is $\la w\ra^2/\Gamma$, where $\sqrt{\Gamma}$ is the strength of the white noise in the Langevin equation (for
the step position) underlying the FP equation \cite{pge}.

To \textcolor{black}{confront} data, both experimental and simulational, one typically investigates the variance $\sigma^{2}(t)$ \textcolor{black}{or standard deviation $\sigma(t)$} of this distribution.
If the initial configuration of the vicinal surface is ``perfect" (i.e.\ has uniformly-spaced straight steps), then $\sigma (t)$ \textcolor{black}{obeys}\cite{pge}

\begin{equation}
\ln\left[\textcolor{black}{1 - }\left(\frac{\sigma\left( t\right)}{\sigma _{\rm sat}}\right)^2  \right] \propto -t/\tau \quad {\rm ;} \quad \sigma _{\rm sat}^2 =  \frac{\left(\varrho + 1 \right)}{2b_{\varrho}}-1
\label{eq:var}
\end{equation}
\noindent where $\sigma _{\rm sat}^{2} \equiv \sigma(\infty)$ is the variance for an
infinite system at long time.  When dealing with numerical data, we take the variance to be normalized by the mean spacing, so divided by the squared mean terrace width, to mimic the formal analysis.  The precise value of the proportionality constant is not of importance to our analysis, since we view $\tau$ as the source of information for an activated process, with any prefactor therefore insignificant.  As discussed in the appendix (esp.\ Eq.~(\ref{eq:sigmu})), one might expect the prefactor to be unity when the first moment has the assumed GWS value of one, but with the approximations we make to obtain a compact solution, the prefactor seems better described as two.

Time in this formulation is not the natural fermionic time associated with the
direction along the steps ($\hat{y}$ in ``Maryland notation"), i.e., that resulting from the standard  mapping between a 2D classical model and a (1+1)D quantum model.  Instead it measures the evolution of the 2D or (1+1)D system toward equilibrium and the thermal fluctuations underlying dynamics.   Since the time constant $\tau$ enters rather obliquely through the noise force of the Langevin equation, a key investigational objective
in the previous Letter\cite{pge} and in this paper is whether $\tau$ corresponds to a physically significant rate.  Monte Carlo
simulations allow the examination of a well-controlled numerical experiment.  In the former we used our well-tested Metropolis algorithm to
study a terrace-step-kink (TSK) model of the surface.  We found a satisfactory fit to the form of Eq.~(\ref{eq:var}), from which we obtained
$\tau\approx$ 714 MCS (Monte Carlo steps per site) for $\varrho = 2$ (or $A = 0$, only entropic repulsions) while $\tau\approx$ 222 MCS for
$\varrho\approx$ 4.47.  This result is in qualitative agreement with the understanding that $\Gamma$ should increase (and, so, $\tau$ should decrease)
with increasing $\varrho$, as discussed in Ref.~\onlinecite{pge}.

In the present paper, we confront more systematically and thoroughly the above-noted crucial issue, showing that the time constant associated with the FP transcription can be related to the atomistic processes underlying the relaxation to equilibrium and that the FP time \textcolor{black}{in some sense} tracks \textcolor{black}{(though of course does not replicate)} the literal physical time of the relaxing system.  We use a standard, simple lattice model that embodies the basic atomistic properties of these surfaces.  We report far more extensive simulations, using kinetic Monte Carlo (KMC)\cite{amar06,Voter} rather than the Metropolis algorithm, for a solid-on-solid (SOS) rather than a TSK model, so that we have real mass transport.  Since atomic energies in this generic model are proportional to the number of lateral nearest neighbors, detailed-balance is satisfied. 
\textcolor{black}{To simplify the analytic expressions and, especially, the simulations, we concentrate in this paper on the special case $\varrho = 2$, corresponding to steps with only entropic repulsions (``free fermions").}
We find that the time constant, extracted from the
numerical data by fitting to the time correlation function in the form predicted by the FP analysis, has an activated form that can
be related to an atomistic rate-limiting process in the simulations.  Our goal is not to find the best accounting for the dynamics of a real stepped surface, nor even of our model surface.  It is to show that the FP approach offers a relatively simple and physically viable approach to accounting for the relaxation of artificial initial configurations toward equilibrium.

The second section describes the model and KMC algorithm that we use.  The third presents our numerical results.  The fourth discusses them, with one subsection describing the crucial role played by the creation of kink-antikink pairs and another investigating the evolution of the shape of the distribution.  The fifth makes comparisons with the venerable mean-field treatment of step distributions, and the final section sums up our findings.  In an Appendix we expand the derivation of the key Fokker-Planck equation given in Ref.~\onlinecite{pge}; we present some new results for the evolving moments of the $P_2(s,\tilde{t})$ and correct some inconsequential errors in Ref.~\onlinecite{pge}.

\section{Model}
Our SOS model assigns an integer height $h_{\bf r}$ to each point {\bf r} on a square grid of dimensions $L_x \times L_y$.   We use periodic boundary conditions in the $\hat{y}$ direction.  On our vicinal (001) simple cubic crystal, we create $N$ close-packed [100] steps, with mean separation $L = L_x/N$, via screw periodic boundary conditions in the $\hat{x}$ direction.  In our simulations we take $N$=5 \textcolor{black}{in the initial simulations\cite{hpe1} and $N$=20 in later investigations}.  The energy of a configuration is given by the standard absolute SOS prescription:

\begin{equation}
{\cal H} = \frac{1}{2} E_a \sum_{\bf r \delta} h_{\bf r} h_{\bf r+\delta}
\label{eq:SOS}
\end{equation}
\noindent where $\delta$ runs over the 4 nearest-neighbors of a site, and the factor 1/2 cancels the double-counting of bonds.

In our SOS model, which has been described elsewhere \cite{Vid}, we use barriers determined
by the \textcolor{black}{standard longstanding simple rule\cite{SVv,J99,CVv} of bond-counting}: the barrier energy $E_b$ is a diffusion barrier $E_d$ plus a
bond energy $E_a$ times the number of lateral nearest neighbors in the initial state.  This number is 1 for an edge atom leaving a straight
segment of step edge for the terrace, 3 for a detaching atom that originally was part of this edge (leaving a notch or kink-antikink pair in the
step), or 2 for a kink atom detaching, either to the step edge or the terrace.  Processes that break 4 bonds, in particular the removal of an atom from a flat terrace plane, are forbidden, as is any form of sublimation.  No Ehrlich-Schwoebel barrier hinders atoms from crossing steps.   We chose values 0.9 $\le E_d \le$ 1.1 eV and 0.3 $\le E_a \le$ 0.4 eV, using temperatures 520K $\le T \le$ 580K.  At these
temperatures we expect no significant finite-size effects in the $\hat{y}$ direction for the values of the mean terrace width $L$ (in lattice spacings) that we use: 4 $\le L \le$ 15.

The width $L_y$ of the lattice should be greater than the ``collision length" $y_{\rm coll}$, the distance along $\hat{y}$ for a step to wander a distance $L$/2 in $\hat{x}$.  For a TSK model, estimates using a random-walk model give $y_{\rm coll} = (L^2/2)\sinh^2(E_k/2k_BT)$ \cite{BEW}, where $E_k$ is the formation energy of a kink.  At the temperatures and energies used in our simulation, $y_{\rm coll}$ is of order $10^2$ for $L$=6 and $10^3$ for $L$=15.  E.g., for $T$=580K, $E_k=E_a/2=0.175$eV, and $L$=6, $y_{\rm coll} \approx 140$.   In almost all simulations reported here, we use $L_y = 10^4$.  While $L_y$ may often be larger than necessary, it allows for some self-averaging, decreasing the number of runs we need to carry out to get good statistics.

In our rejection-free KMC, we separate all top-layer sites into 4 classes, those with $i$=0,1,2,3 nearest neighbors (NNs).  (Those with $i$=4 are not allowed to move and are not considered when updating.)  Typical realizations of these four classes are isolated adatoms, atoms protruding from a straight step edge, atoms at kink sites, and atoms at the edge of a step, respectively.
We compute probabilities for each of the movable classes: $P_i = f_i/ \sum_{i=0}^3 f_i$, where
$f_i = N_i \cdot \exp[-(E_d + i*E_a)/k_BT]$, and $N_i$ is the number of sites with $i$ NNs.  (Of course, the 4 exponentiations are done once and for all at the beginning \textcolor{black}{for each set of energies.})  For each update we need four random numbers---$r_1,r_2,r_3,r_4$---uniformly distributed between 0 and 1.  We use $r_1$ to pick which of the 4 movable classes will have the move.
For the ``winning" class, $r_2$ determines which of the $N_i$ possible atoms will move.
Then $r_3$ determines in which of the 4 NN directions the atom moves.
In this rejection-free scheme, we then decrease the height (the z value) of the initial position by one and increase the height of the chosen direction move from this initial site by one.
This scheme can be (and has been, elsewhere) modified to allow for an Ehrlich-Schwoebel barrier.
Finally $r_4$ is used to advance the clock in standard KMC fashion, similar to the n-fold way or
BKL\cite{BKL} approach, using the prescription $\Delta t = - \ln(r_4)/R$, where $R$ is the
total rate for a transition from the initial state \cite{amar06}.  Explicitly, $R = 
\nu_0 \sum_{i=0}^3 f_i =
\nu_0  \exp[-E_d/k_BT]\sum_{i=0}^3 N_i \cdot \exp[-i*E_a/k_BT]$, where we
take \textcolor{black}{the hopping frequency} $\nu_0 = 10^{13}$ s$^{-1}$.  

 We saved essentially every hundredth update; that interval  corresponds to our unit of time, which is about 1~sec.\ for the selected temperature and energies \cite{time}.   This update interval is long enough so that the sum of the KMC update times varies insignificantly ($\pm 0.01$\%)
but short enough to capture the behavior during the steep initial rise.

In our model, the mass carriers are atoms rather than vacancies (or both).  
Since atoms with $i$=4 are frozen in our model, 
atom-vacancy pairs cannot form spontaneously on a terrace.  (More generally, this mechanism is highly improbable.)  Mass carriers are thus created at step edges.  If MC moves depend on the difference between final and initial energies, as in Metropolis schemes, then there is equivalence between atom and vacancy creation and transport.  (If one goes beyond a strict SOS model and allows local relaxation, vacancies tend to be favored somewhat \cite{nelson,vac}.)  An atom quitting a step edge for the lower terrace costs 3$E_a$ if it leaves a straight step and 2$E_a$ if it leaves from a kink.  At the upper side of a step, a vacancy can be spawned if a step-edge atom moves out one spacing onto the lower terrace (with the same energy cost as just given) and its inner neighbor happens to move in the same direction before the initial atom returns to its initial position.  
In kinetic Monte Carlo, however, rates are determined just by the difference between the barrier energy and the initial-state energy.  This does not change the energy to produce an atom, but adds a cost of 3$E_a$ for the move of the second, inner-neighbor atom.  Moreover, while the energy for an atom to hop along the terrace is $E_d$, for a vacancy it is $E_d + 3E_a$.  
Indeed, we never observed the unlikely concerted process for vacancy creation in our simulations nor, for that matter, did we see any vacancies.  The number of isolated atoms was also very small, with $N_0$ being in single digits, and they moved very rapidly, rarely appearing in successive saved images.   

The freezing of $i$=4 processes marks a violation of detailed balance (since such a vacancy, if it existed, could be filled by a roving adatom); however, given the negligible occurrence of such vacancies in our simulations, the violation should be insignificant.  In some physical systems, motion of surface vacancies does evidently dominate mass transport \cite{vGastel}.  Again, our goal in these calculations is not to account generally for experiments but to create a fully-controlled data set to see how well the dynamics can be described using our Fokker-Planck formalism.

\section{Computed Results}
\begin{figure}[t]
\includegraphics[width=8 cm,clip=true,trim=0mm 6mm 0mm 6mm]{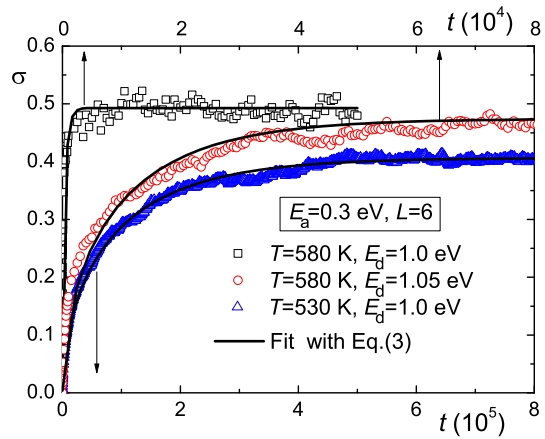}
\caption {Three examples of fits using Eq.~(\ref{eq:var}), used to extract $\tau$.  Note that the data are very well fit in all three cases. The plotted $\sigma(t)$ is the standard deviation of the KMC data divided by the mean step spacing $L$ (listed in lattice constants), and $E_d$ and $E_a$ are the energy barriers for diffusion and for breaking a bond, respectively. Time $t$ is essentially in seconds (see text).  \textcolor{black}{In all cases here and in later figures, $\varrho$=2.} \textcolor{black}{Here $N$=5.}
}
{\label{f:fit}}
\end{figure}

We extract a \textcolor{black}{characteristic} time (or inverse rate) $\tau$ from numerical data by fitting the dimensionless width using
Eq.~(\ref{eq:var})\textcolor{black}{, as illustrated in Fig.~\ref{f:fit}}.  
The fit is notably better than that found in the Metropolis/TSK study in Ref.~\onlinecite{pge}.
[However, the saturation value is notably higher than in Ref.~\onlinecite{pge}, with the normalized standard deviation $\sigma$ (the value in the simulation divided by $L$) approaching $\sim$0.48, or a dimensionless variance of 0.24, rather than 0.18 as found in Ref.~\onlinecite{pge} and anticipated from Eq.~(\ref{eq:var}).  This difference arises because the present algorithm allows steps to make contact along edge links rather than just at corners as in the usual fermion simulations.  The variance of  0.18 is appropriate to ``free fermions" with $\varrho$=2.  As we discuss in detail elsewhere \cite{kiss}, the present algorithm leads to a smaller (and $L$-dependent) effective $\varrho$  as the steps come in contact more frequently, i.e. for smaller $L$ and higher $T$ (cf.\ Fig.~\ref{f:fit}).  For the present choice of parameters ($L$=6, $k_BT/E_d \! \approx \!$ 1/20), the TWD has close to $\varrho$=1, for which the dimensionless variance is 0.27.  This feature is inconsequential for the arguments in this paper.]

\begin{figure}[t]
\includegraphics[width=8.2 cm,clip=true,trim=6mm 6mm 10mm 4mm]{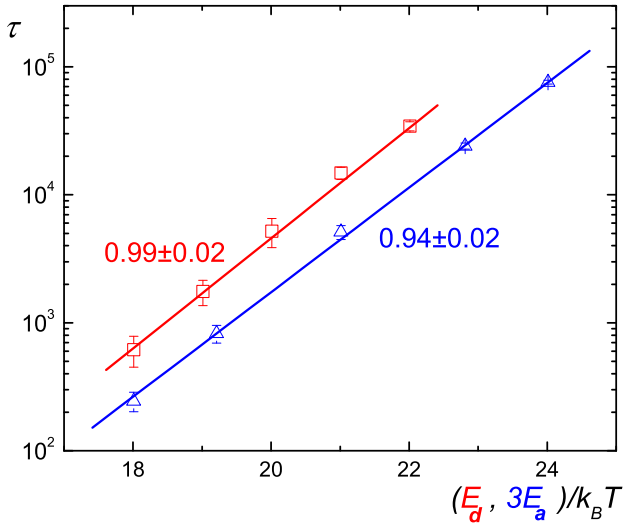}
\caption {Semilog plots of the relaxation time $\tau$ (in sec.) vs.\ the diffusion barrier $E_d$ \textcolor{black}{(squares, upper line, red) or thrice the bond energy $E_a$ (triangles, blue), with the other held fixed, both in eV, with $k_BT\! =\! 0.05$eV and $N$=5.  The numbers indicate the slopes; both are essentially unity.}
}
\label{f:ramp}
\end{figure}

\begin{figure}[t]
\includegraphics[width=\columnwidth,clip=true,trim=5mm 6mm 12mm 6mm]{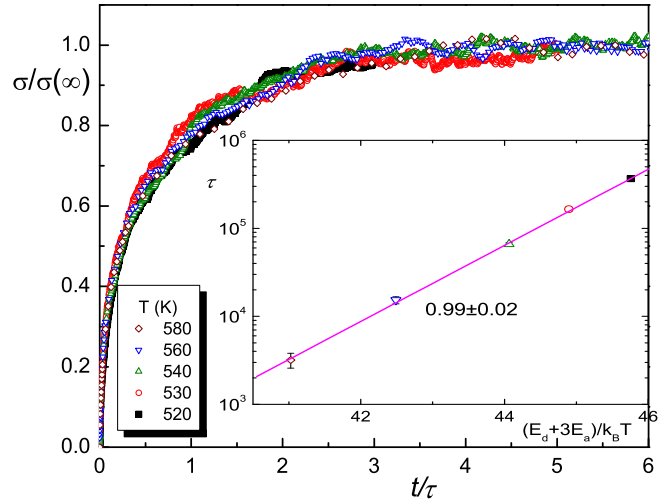}
\caption {Demonstration of the robustness of the form of the evolving standard deviation, $\sigma(t)=\sigma(\infty)\left[1-\exp(-t/\tau)\right]^{1/2}$, for five temperatures.   In the inset, the five values of $\tau$ by which the curves are rescaled are plotted vs.\ \textcolor{black}{$(E_d +3E_a)/k_BT$}, showing their Arrhenius form.  Here $E_d$=1.0eV and $E_a$=0.35eV. \textcolor{black}{In this and subsequent figures, $N$=20.}
}
\label{f:SigTauT}
\end{figure}

\textcolor{black}{We expect that the decay time exhibits Arrhenius behavior:} $\tau \propto \exp(E_b/k_BT)$. We investigate $E_b$ closely in the two traces of
Fig.~\ref{f:ramp}. We show typical runs at $T$ = 580K, corresponding to $k_BT \approx$ 1/20 eV.  First, we ramped $E_d$, holding
$E_a$ fixed at 0.35 eV (open squares, red). In the semi-log plot of reduced energies (energies/$k_BT$), we find a  slope of 0.99 $\pm$ 0.02, indicating that in the
effective barrier, the multiplier of \textcolor{black}{$E_d/k_BT$, is essentially} unity, as expected.  In a second set of runs, we ramped $E_a$, holding $E_d$ fixed at 1.0 eV (open
triangles, blue).  Plotting now vs.\ $3E_a/k_BT$, we find a slope 0.94 $\pm$ 0.02, 
indicating that the effective energy barrier $E_b$ is $E_d + 3 E_a$.  
To \textcolor{black}{corroborate} this idea, we ramped \textcolor{black}{$T$}
from 520 to 580K, fixing $E_a\! =\! 0.35$eV and $E_d\! =\! 1.0$eV. As \textcolor{black}{illustrated} in the inset of Fig.~\ref{f:SigTauT}, we \textcolor{black}{determine the fitted activation energy to be} 2.03 $\pm$ 0.03, in excellent agreement with $E_d + 3E_a$ = 2.05 [eV]. Evidently the rate-determining process is the removal of a 3-bonded atom from a straight step, creating a pair of kinks (i.e., a kink and an antikink\cite{VF}) rather than the presumably more frequent process, with energy $E_d + 2 E_a$, in which an atom leaves a kink position of a step \cite{VF,G01}.  \textcolor{black}{(Of course, kink-antikink pairs also arise with a lower barrier when an atom from the terrace or from a kink site attaches to a step edge or splits off from a kink site.  However, as members of class $i=1$, such edge-atom structures are likely to be very short-lived.)}  The main part of Fig.~\ref{f:SigTauT} shows the standard deviation ($\propto$ TWD width) vs.\ time scaled by the relaxation time of each of the five temperatures.  Evidently the fit to $\sigma(t)=\sigma(\infty)\left[1-\exp(-t/\tau)\right]^{1/2}$ \textcolor{black}{is} robust. 

\begin{figure}[t]
\includegraphics[width=4cm]{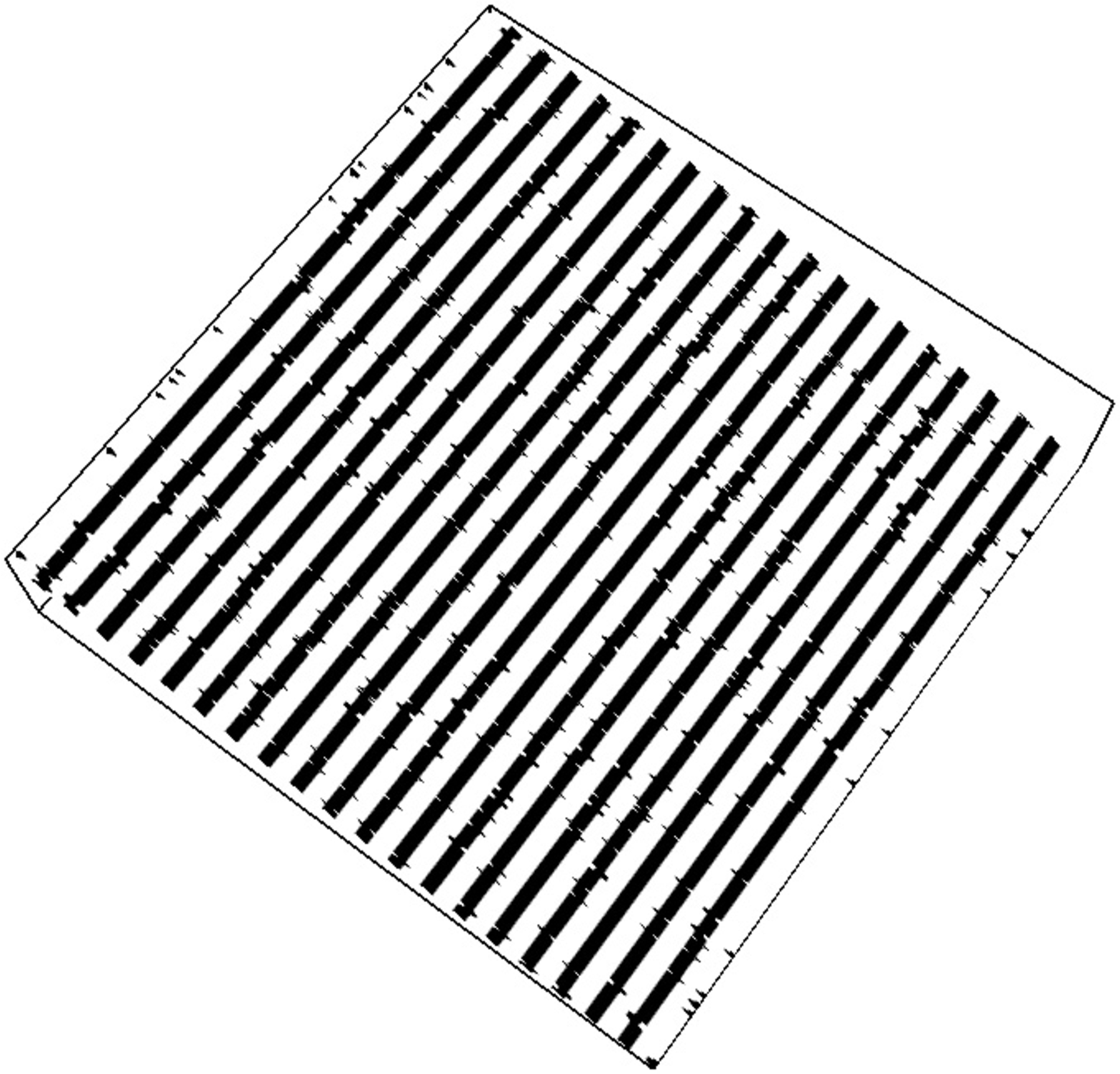}
\includegraphics[width=4cm]{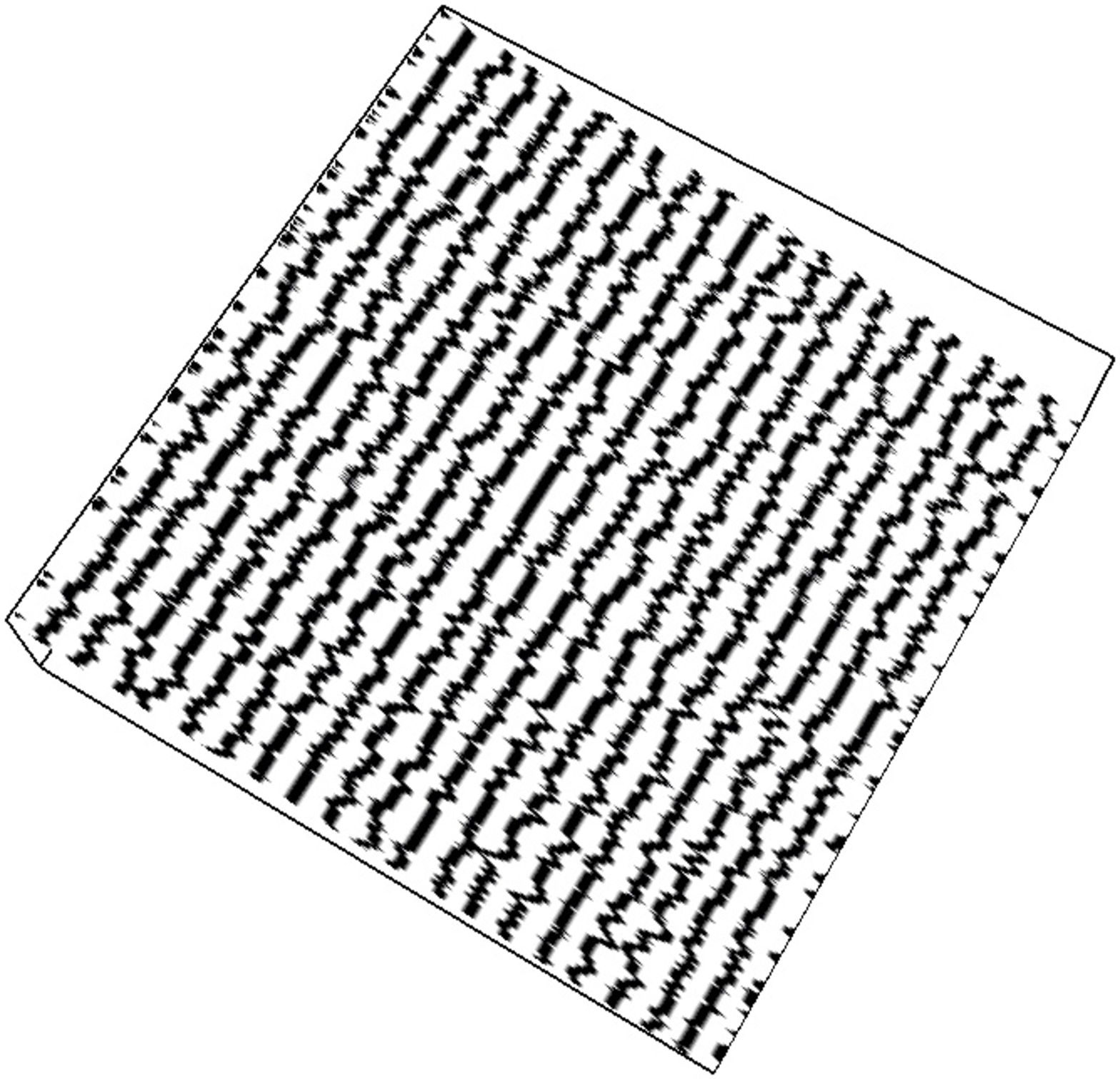}
\includegraphics[width=4cm]{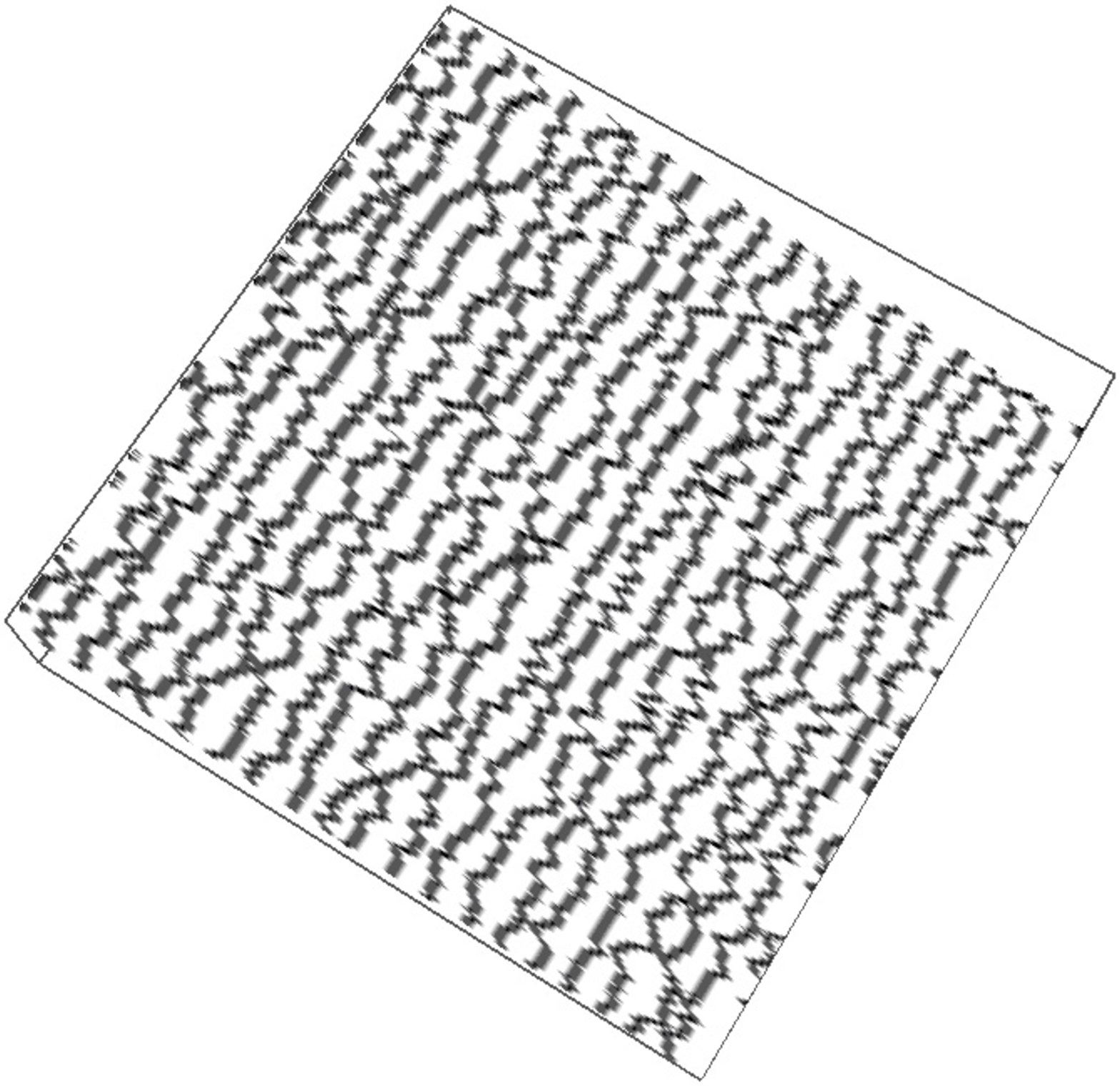}
\includegraphics[width=4cm]{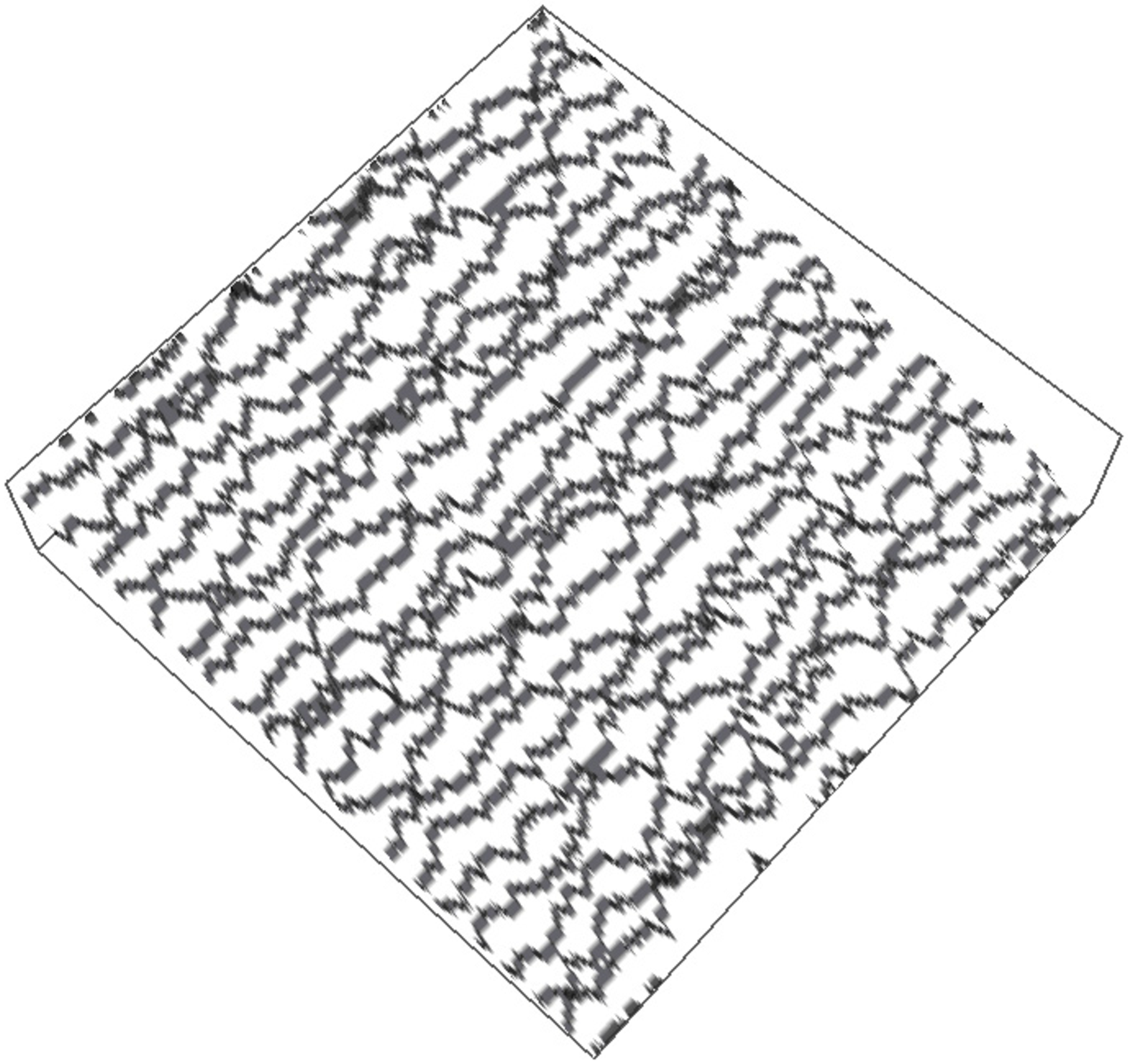}
\caption {\textcolor{black}{Typical step configurations during the evolution of initially straight steps in Fig.~\ref{f:SigTauT}.  The panels are 120 $\times$ 5000 site portions extracted from the full 120 $\times$ 10000 net; there is considerable compression in the $\hat{y}$ direction.  The panels range from early time to near saturation.  Specifically, the ratios of the image time to $\tau$ are: $\sim 1/80$, $\sim 1/8$, $\sim 1/2$, and somewhat over 3.  From the step images alone, one would be hard pressed to distinguish the uphill direction, which is to the left. 
}}
\label{f:flucsym}
\end{figure}

\textcolor{black}{Initially the steps retreat as atoms are emitted. There is also an asymmetry in fluctuations from a straight step, since retreating moves involve higher barriers than advancing fluctuations.  Once the continuum picture becomes applicable, the fluctuations appear to be symmetric,} \textcolor{black}{with typical configurations shown in Fig.~\ref{f:flucsym}.}

\textcolor{black}{To check} consistency, we compare the intercepts of the linear fits in the two semilog plots, i.e., the prefactors of the exponential term in which the \textcolor{black}{particular} energy is ramped.  In addition to the activation components there is the leading factor
$\tau_0 \equiv \la w \ra^2/4 \nu_0$\textcolor{black}{ \cite{pge,nu4}, where we make the standard assignment for the hopping frequency,} $\nu_0\! = \!10^{13}$Hz.  Since $\la w \ra \! \equiv \! L \! = \! 6$ in our simulations, $\tau_0$, theoretically expected to be $9\times10^{-13}$s, is found in the simulations to be $(8.71\pm 0.25)\times10^{-13}$s.  In the ramp of $E_d$, the prefactor is $\tau_0 \exp(3E_a/k_BT)$, predicted to be $1.19\times10^{-3}$s.  The value we find from the simulations is $(1.2\pm 0.1)\times10^{-3}$s, in excellent agreement.  Similarly in the ramp of $E_a$, the prefactor $\tau_0 \exp(E_d/k_BT)$ is predicted to be $4.366\times10^{-4}$s and measured from the fit as  $(4.36\pm 0.15)\times10^{-4}$s.

We also varied the system size \textcolor{black}{$L_x$}, holding the number of steps fixed, and thereby ramping $\langle w \rangle$. From the random-walk analogy, the
prediction is that $\tau \propto \langle w \rangle^2$.  We find tolerable agreement, with a slope 18\% below the expected value.   We suspect that the reason behind the poorer agreement than for $L$=6 above originates in the $L$-dependence of the variance associated with the peculiar algorithm used in our simulations, which allows steps to touch \cite{kiss}.

\textcolor{black}{Independently, another} argument corroborates that the kink creation rate has an
activation energy of $E_d+3E_a$: \textcolor{black}{At} equilibrium the creation and the annihilation rates are equal, \textcolor{black}{so} we compute the latter. Annihilation of kinks requires that an adatom diffuses to a \textcolor{black}{step-edge} notch---a kink-antikink pair, whose density is $n_{k-ak}\approx\exp{(-2E_k/k_BT)}\approx\exp{(-2(E_a/2)/k_BT)}$.  Since the equilibrium adatom density is $c_{\rm eq}=\exp(-2E_a/k_BT)$, the annihilation rate of kinks at a step edge is proportional to $Dc_{\rm eq}n_{k-ak}\sim
\exp{[-(E_d+2E_a+E_a)/k_BT]}$ (cf. Ref.~\onlinecite{KK03}). This in turn implies that the activation energy for the kink creation rate is  $E_d+3E_a$.

\section{Discussion of Results}
\subsection{Crucial role of kink creation}

\textcolor{black}{T}he key energy in the relaxation time is that for detaching 3-bonded atoms rather than kink atoms, \textcolor{black}{a remarkable observation.}  Neither equilibrium nor growth
processes involve 3-bonded atoms. At equilibrium, step fluctuations are controlled by the so-called step mobility, which is proportional to the emission rate
of adatom from kinks \cite{PV}, \textcolor{black}{which} involves 2-bonded atoms. \textcolor{black}{Hence,} the relaxation towards equilibrium \textcolor{black}{should also be} controlled by the
step mobility. However, our results \textcolor{black}{evidently contradict this notion.}

\begin{figure}[t]
\includegraphics[width=8 cm,clip=true,trim=6mm 2mm 6mm 6mm]{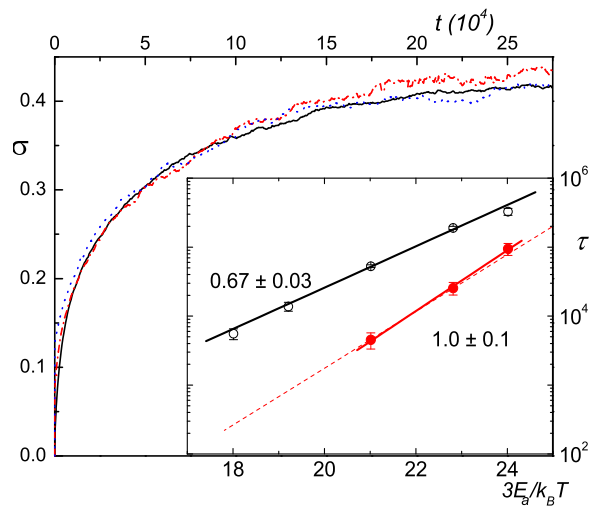}
\caption [Checks]{Checks of dependencies on initial conditions and update moves.  Evolution of the standard deviation $\sigma$ of the TWD for three initial configurations: straight steps (solid, black), ``decimated" edge (dash-dotted, red), and crenelated (dotted, blue) edge.  The 120 $\times$
10,000 lattice has 20 steps, with $L$=6; we choose $T$=580K, $E_d$=1eV, and $E_a$=0.4eV.  For equilibrated non-interacting (free-fermion-like, $\varrho \! = \! 2$) steps, $\sigma$ approaches 0.42 $= \sigma_W$, as expected. The smooth curve is a fit to an exponential approach to saturation.  The smooth curve is $\sigma(t) = \sigma(\infty) \left[1-\exp(-t/74852)\right]^{1/2}$. {\it Inset}: Surface azimuthally misoriented by 0.0005 radians, forcing 5 kinks along the $10^4$-site steps.  The three filled (red) circles represent runs with an initial ``perfect" configuration of 5 straight 2000-site segments; the slope, indicated by the solid red line, is 1.0$\pm$0.1, in excellent agreement with the steeper line in Fig.~2.  Thus, processes in which 3 bonds are broken govern the scaling of the relaxation time $\tau$ (given in sec. as in Fig.~\ref{f:fit}). The open circles are from runs with 3-bonded atoms immobile; the corresponding slope is 0.67$\pm$0.03, so that now 2-bonded atoms control the (much larger) $\tau$. 
}
\label{f:kink}
\end{figure}

One \textcolor{black}{possible} explanation \textcolor{black}{of our remarkable} finding is that kinks have to be formed
first, requiring the extraction  of atoms from straight \textcolor{black}{steps. (As noted earlier, the addition of an atom to a step edge also creates kink-antikink pairs, but the ``tooth" is of class $N_1$, so very short-lived.)}. In that case, our initial configuration, in which steps are perfectly parallel and straight may be introducing a bias in the results.  To investigate this possibility, we considered two other initial states with equal numbers of kinks and antikinks, i.e. in which one produces kinks by adding atoms to (or removing atoms from) a straight edge \cite{polarmis}.  In the ``decimated" case every tenth atom along a straight step is removed; in the other case every other atom is removed to create a ``fully kinked" step, so that the edge resembles dentil molding or castle crenelations.    In Fig.~\ref{f:kink} we plot the resulting evolution of the standard deviation $\sigma$ of the TWD for the three cases.   During the early-time rapid spreading of the initial sharp TWD, the slope increases with the number of initial kinks. In this regime, our model is not expected to apply (nor should any other 1D, continuous model); indeed, Eq.~(\ref{eq:var}) does not describe the steep initial part of the traces very well.  After about $4\! \times \! 10^4$ MCS the curves are essentially indistinguishable within the noise level. The smooth curve is a one-parameter best fit of the data for the initially straight step by $\sigma(t)/\sigma(\infty) = 1-\exp(-t/\tau)$. For this case we find $\tau \approx 7.5\!  \cdot \! 10^4$ (in units that are essentially sec.).

We formulated the crenelated configuration because it creates at the outset a high density of atoms of class $N_1$: $8\times 10^{-2}$ atoms/site, three orders of magnitude greater than the equilibrium density of $5\times 10^{-5}$.  These atoms quickly lead to a burst of adatoms that should quickly thermalize the step configurations.  If it were only the supply of adatoms that limits equilibration, then for this scenario the subsequent creation of kinks would not be crucial.   Evidently this is not so; even for the crenelated case, the relaxation kinetics are determined by the rate of creating kinks-antikinks pairs, with the usual energy barrier. 

To corroborate that kink creation is indeed the rate-limiting process, we computed the relaxation rate of a surface with steps azimuthally misoriented so as to create kinks via screw boundary conditions in the $\hat{y}$ direction. Specifically, in the
initial state the in-plane misorientation slope was set at 0.0005, so that geometry forces the existence of 5 kinks for $L_y$=10,000. Keeping the diffusion barrier fixed at 1 eV, we varied $E_a$. The results are shown in the inset of Fig.~\ref{f:kink} as filled circles. We computed just three points, but clearly, essentially no difference is found with respect to the relaxation rate of straight [100] steps (the red line in Fig.~\ref{f:ramp}). The latter
is drawn as a dashed line in the inset. The fitted slope to the data (times $k_BT$) is 3.0 $\pm$ 0.3, fully consistent with 3-bonded ledge
atoms being responsible for the rate-limiting process.  We also checked that the relaxation rate is enormously slowed if 3-bonded atoms are kept immobile:  Fitting the distribution width with Eq.~(\ref{eq:var}), we  ramped $E_a$ while holding fixed $E_d =1$ eV. The extracted relaxation times are shown in the inset of Fig.~\ref{f:kink} as open circles. The reduced slope is 2.0 $\pm$ 0.1, consistent with 2-bonded kink atoms providing the rate-limiting process for the step motion in this case.  The characteristic time is at least an order of magnitude larger than the previous case, which can be interpreted as due to the inability to create new kink sites, so that the number of sources for 2-bond escape of atoms to the straight segments of the step is limited to the initial 5 kinks.  Without the azimuthal misorientation, this surface would be inert.  Furthermore, the eventual width of the distribution, $\sigma _{\rm sat}$, is only about half the size of the 3-bond case.  Thus, at least over the course of our long runs, the surface is never able to equilibrate.


We considered the number of $N_2$ sites, typically kinks along steps.  This quantity rose much more rapidly than the variance.  Referring to the main plot of Fig.~\ref{f:kink}, $N_2$ achieves its saturation value by $t \approx 3\! \times \! 10^4$.  Thus, the process controlling $\tau$ is not the initial formation of an adequate number of kinks but rather the maintenance of this number.  For our chosen energies, about 1 in 30 sites along a step was a kink, far higher than in our azimuthally slightly-misoriented case.

We also applied a similar analysis of the variance of the TWD to a vicinal (001) surface misoriented in along an azimuth rotated 45$^\circ$ so as to have zig-zag [110] steps.  For such steps, every outer atom has $i$=2 lateral neighbors.  Our analysis then shows that these 2-bond kink atoms produce the rate-limiting step, with a slope of 2 in the equivalent of the plot of $\tau$ vs.\ $E_a$ in the inset of Fig.~\ref{f:kink}.  This system has some idiosyncratic behavior due to the ease of creating fluctuations of steps from their mean configuration.  Discussions of these subtleties would cloud the focus of this paper. Hence, we defer details to a future communication \cite{kiss}.

In short, we reach the striking conclusion that the equilibration of a terrace width on a vicinal (001) simple cubic crystal with close-packed [100] steps (or steps not far from close-packed) and the fluctuations of the same terrace width at equilibrium are qualitatively different phenomena.  The latter can take place with a constant number of kinks, while the former requires creation of new kinks.  Fluctuations from the equilibrium distribution, having the form of Eq.~(2), will not lead to arbitrary initial configurations such as a perfect cleaved crystal with straight, uniformly-spaced steps.

\subsection{Higher moments of the TWD}
\label{s:HiMo}

\begin{figure}[t]
\includegraphics[width=8 cm,clip=true,trim=6mm 4mm 2mm 7mm]{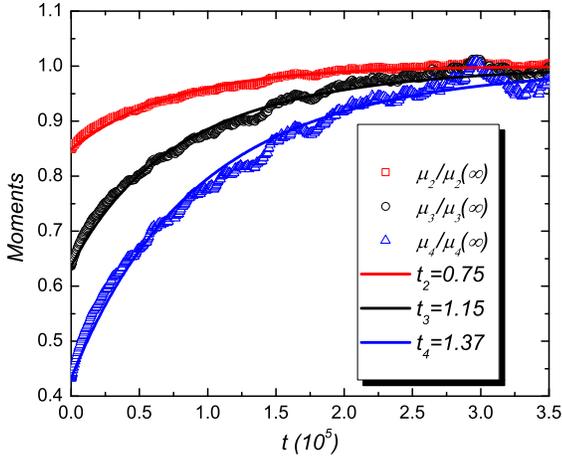}
\caption [Fig45]{From top to bottom, the second (squares, red), third (circles, black), and fourth (triangles, blue) moments (with respect to the origin) of the evolving KMC-generated TWDs for initially straight configurations and the same parameters as in Fig.~\ref{f:kink}; for ease of comparison, the $j^{\rm th}$ moment $\mu_j(\tilde{t})$ is divided by its equilibrium value $\mu_j(\infty)$.  To make contact with the exponential approach of these moments to their saturation values, one must rescale the KMC time.  For $\mu_2$ the rescaling factor, as in Fig.~\ref{f:kink}, is $7.5 \! \cdot \! 10^4$.  For $\mu_3$ and $\mu_4$, the approach to saturation is progressively slower; the rescaling times are $1.15 \! \cdot \! 10^5$ and $1.37 \! \cdot \! 10^5$, 3/2 and 9/5 as large, respectively. 
}
\label{f:mu2340}
\end{figure}

\begin{figure}[t]
\includegraphics[width=8.5 cm,clip=true,trim=10mm 2mm 4mm 2mm]{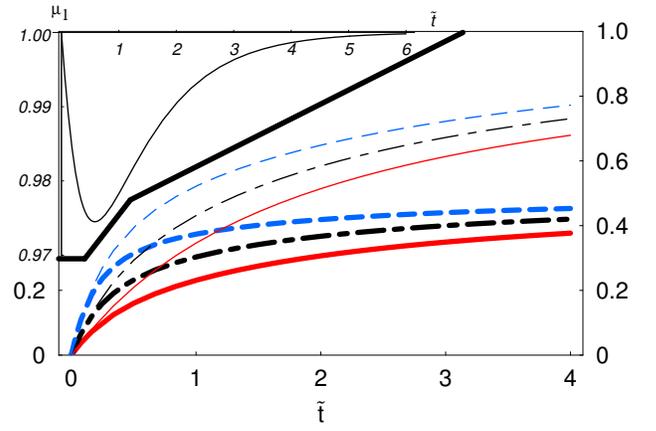}
\caption {Analytic results for the moments of the evolving TWD predicted by the (1D continuum) Fokker-Planck theory:  Plot of the effective exponential decay time of the difference between the evolving second (solid, red), third (long-short-dashed, black), and fourth (dashed, blue) moments of the  solution to Eq.~(\ref{eq:fp3}), $P_{\varrho\! =\! 2}(s,\tilde{t})$, given explicitly in the appendix in Eqs.~(\ref{eq:mu2}), (\ref{eq:mu3}), and (\ref{eq:mu4}), and their steady-state, asymptotic values associated with Eq.~(\ref{eq:Ps}).  As in the KMC data in Fig.~\ref{f:mu2340}, the decay is significantly slower for the higher moments.  The thicker set of curves corrects for the modest deficiency of the the first moment $\mu_1(\tilde{t})$; $\mu_1(\tilde{t})$ is depicted in the inset (upper left, italicized axes labels) and given analytically in Eq.~(\ref{eq:mu1}).  See text for details. 
}
\label{f:tau-mu1}
\end{figure}

So far, we have only considered the first two moments of the TWD.  Several different distributions might account for these two moments.  As a further check that $P_2(s,\tilde{t})$ describes the KMC data well, we study higher moments of the TWD in comparison with the analytical expressions in Eqs.~(\ref{eq:mu3}) and (\ref{eq:mu4}).  In Fig.~\ref{f:mu2340} we plot the second, third, and fourth moments (with respect to the origin) for initially-straight steps. (Since we wish to compare with $P_2(s,\tilde{t})$, we divide the ``raw" KMC $j^{\rm th}$ moment by the $j^{\rm th}$ power of the mean spacing to determine $\mu_j(\tilde{t})$.)  For each moment  there is a steady rise (from unity) that approaches the equilibrium value $\mu_j(\infty)$ exponentially.  (Steps which are initially decimated or crenelated behave similarly, though in somewhat ``noisier" fashion.) For $\mu_2$, $\mu_3$, and $\mu_4$, these saturation values agree well with the analytic results $3\pi/8 \approx 1.18$, $\pi/2 \approx 1.57$, $15\pi^2/64 \approx 2.3$, respectively, as which can be read off Eqs.~(\ref{eq:mu2}), (\ref{eq:mu3}), and (\ref{eq:mu4}).  For ease of comparison, we plot $\mu_j(\tilde{t})/\mu_j(\infty)$ in Fig.~\ref{f:mu2340}, so that each normalized moment approaches unity. The approach to saturation is evidently slower for successively higher moments.  

To make contact between the moments extracted from the KMC data and the analytic expressions in dimensionless units arising from our Fokker-Planck analysis, we seek whether by rescaling KMC times by some $t_j$ leads to a good description of the data by the deduced moments.  For $\mu_2$ such a rescaling factor $t_2$, with  $t_2 = 7.5 \! \cdot \! 10^4$, was already used in analyzing the data in Fig.~\ref{f:kink}.  In other words, we adjust $t_2$ so that $\mu_2(t/t_2)/\mu_2(\infty)$ from Eq.~(\ref{eq:mu2}) fits the data as well as possible, as illustrated in Fig.~\ref{f:mu2340}.  For $\mu_3$ and $\mu_4$, the approach to saturation is progressively slower; the rescaling times $t_3$ and $t_4$, similarly obtained, are $1.15 \! \cdot \! 10^5$ and $1.37 \! \cdot \! 10^5$, 3/2 and 9/5 as large, respectively.

The analytic expressions for the four moments, given in Eqs.~(\ref{eq:mu2}), (\ref{eq:mu1}), (\ref{eq:mu3}), and (\ref{eq:mu4}), all approach saturation asymptotically from below like $\exp(-\tilde{t})$.  However, in the temporal regime corresponding to the KMC simulations, higher-order terms cause the evident exponential-like approach to depend on $\tilde{t}/\tau$ rather than simply $\tilde{t}$, where $\tau$ is some effective time constant of order unity.

To determine $\tau_j$ we consider in Fig.~\ref{f:tau-mu1} the evolution of $-\tilde{t}/\ln[1\! -\!  \mu_j(\tilde{t})/\mu_j(\infty)]$ for each moment ($j$=2,3,4).  To the degree that this trace is horizontal, the Ansatz is appropriate.  The thin curves in Fig.~\ref{f:tau-mu1} show that this assumption becomes progressively better as time advances.   While all three curves eventually converge to unity, in the time regime under consideration the three moments have significantly different time constants, with the higher moments having progressively larger magnitudes, consistent with the KMC findings displayed in Fig.~\ref{f:mu2340}.  

The inset of Fig.~\ref{f:tau-mu1} displays the first moment of $\mu_1(\tilde{t})$.  As described in the appendix, it is a couple percent smaller than the proper value of unity in the region around $\tilde{t} \approx 1$.  (This is clearly a deficiency only of the analytic results.  The KMC data have $\mu_1(\tilde{t}) \equiv 1$ by construction.)  Analogous to the transformation of the TWD from a function of $w$ to $P(s,\tilde{t})$, where $s \equiv w/\langle w \rangle$, we can rewrite the distribution in terms of $s/\mu_1(\tilde{t})$ and show that then the ``corrected" moments $\mu_j^{\rm corr}(\tilde{t}) = \mu_j(\tilde{t})/\mu_1^j(\tilde{t})$.  Obviously, $\mu_1^{\rm corr}(\tilde{t}) \equiv 1$.  The higher moments $\mu_2^{\rm corr}(\tilde{t})$, $\mu_3^{\rm corr}(\tilde{t})$, and $\mu_4^{\rm corr}(\tilde{t})$ are displayed as the thick curves in Fig.~\ref{f:tau-mu1}.  These curves flatten considerably sooner than the thin curves, and to a value $\sim 1/2$, reminiscent of Eq.~(\ref{eq:sigmu}).  For these curves, $(\tau_3 - \tau_2)/\tau_2$ is somewhat over 0.1 while  $(\tau_4 - \tau_2)/\tau_2$ is about twice as large, capturing the trend of the numerical data.  If we use the analytic expressions for the corrected moments as the rescaling factors, then the time rescaling factors are $1.75 \! \cdot \! 10^5$, $1.9 \! \cdot \! 10^5$, and $2.0 \! \cdot \! 10^5$, respectively.  Then  $(\tau_3 - \tau_2)/\tau_2 \approx 0.09$ and $(\tau_4 - \tau_2)/\tau_2 \approx 0.14$, closer to the KMC values. 

Accounting for the skewness and the kurtosis poses a more difficult test for our kinetic Monte Carlo simulation of our model.
Due to the interplay of several moments in computing these statistics, our numerical data is not adequate to test these predicted behaviors meaningfully.
Such analyses are arguably the most stringent tests, in which we seek differences from random-walk, Gaussian behavior; they corroborate the conclusion above that the surface never fully equilibrates.  Far more extensive computations might clarify this matter, but are beyond the scope of our present analysis.

\section{Comparison with Gruber-Mullins}
To help place our approach in context, we compare it with previous approximations in the literature based on the fermion description of the fluctuating steps. The celebrated Gruber-Mullins (GM) approximation \cite{GM} considers a fluctuating
step between two fixed neighbors treated as rigid boundaries. In fermion language, the step is the
1D trajectory of a quantum particle confined to the segment $(0,2\langle w \rangle)$ by an infinite potential. Two cases are easily treated: For non-interacting
steps, the fluctuating step is then the trajectory of a free fermion, and is equivalent to a classical particle performing a random walk in a potential of the form\cite{Risken}
\begin{equation}
V(x)=-{2}\ln[\sin(\pi x/2\langle w \rangle)].
\end{equation}
\noindent Then $s \equiv w/\langle w \rangle$ obeys the Langevin equation
\begin{equation}
\dot s={\pi\gamma\over\langle w \rangle^2} {1\over\tan(\pi s/2)}+\eta.
\end{equation}
This approximation preserves the logarithmic behavior of the repulsive potential at short range; even the amplitude is correct: $(2b_\varrho s -\varrho/s)|_{\varrho=2} = (8s/\pi -2/s)$ in Eq.~(1) is replaced by $-\pi \cot(\pi s/2)$, nearly the same for $s<0.7$. However, the GM potential has bogus
symmetry about $\langle w \rangle$, truncating the long-range tail of $P(s)$.

For strongly interacting steps,
the fluctuating step feels an (approximately) quadratic confining potential, and the TWD distribution is predicted to be Gaussian. \cite{GM}. In our formalism the replacement is now $(s \! - \! 1)/\sigma_G^2$, where $\sigma_G^2$ is the variance of the Gaussian TWD (so half the variance of the associated ground-state wavefunction, which we used to construct the FP potential \cite{Risken}.) This replacement should be compared with $(2b_\varrho s  \! - \! \varrho/s) \!  \approx  \! (\varrho + \frac{1}{2}) s -\varrho/s$ \cite{GE}.  In the GM approximation, $\sigma_G^{-2} = \left[12\varrho (\varrho -2)\right]^{1/2}$, with $(12)^{1/2} \! \approx  3.5$ replaced by $(2\pi^4/15)^{1/2} \!  \approx  3.6$ if the interactions with all steps rather than just the two bounding steps are considered \cite{GE}.  An improved approximation (``modified Grenoble") gives $\sigma_G^{-2} \! \approx 2.1\varrho$ \cite{GE}. Our expression for the FP potential now differs at small $s$ from that derived for these approximations because the Gaussians actually extend (unphysically, albeit with insignificant amplitude) to negative values of $s$.  Our
approach is globally superior to the celebrated GM approximation (as well as to the usual alternatives\cite{tle}), both quantitatively and qualitatively, for all physical values of the step-step interaction strength \cite{Hailu}.
Moreover, our FP equation (\ref{eq:fp3}) is fully soluble, so that the TWD can be obtained analytically as a function of time.

\section{Summary}
In summary, we have shown that the relaxation time of the variance of the solution of our Fokker-Planck equation \textcolor{black}{for step relaxation on a vicinal surface can be fit to the comparable variance in a kinetic Monte Carlo simulation of the standard simple model of atomistic processes at surfaces.  This time has Arrhenius behavior that is related to microscopic processes, substantiating that this FP approach can offer useful physical insight into the evolution of complex surface structures toward equilibrium.  Thus, once the continuum formalism becomes appropriate, the FP time in some sense tracks actual time in our model of an evolving physical system of steps with no energetic repulsion.  The formalism also readily allows such repulsions, inviting future simulations to test how well the Fokker-Planck formalism describes such systems.  Since the steps communicate from the outset, the continuum formalism might apply sooner.  For the situation we have considered, we have argued in several ways that the rate-determining \textcolor{black}{process in step relaxation} is the creation of kink-antikink pairs.  We have also examined higher moments of the distribution, both analytically and with simulations.  While we make no pretense that our approach is either exact or a formal theory, we have shown that it can be a fruitful way to treat relaxation of steps on surfaces.  Many avenues of extension are possible.}

\section*{Acknowledgments}
Work at U. of Maryland was supported by the NSF-MRSEC, Grant DMR 05-20471; visits
by A.P.\ supported by a CNRS Travel Grant. T.L.E.\
acknowledges the hospitality of LASMEA at U.\ Clermont-2. We acknowledge the insightful collaboration of Hailu Gebremariam in the work discussed in the appendix, helpful
conversations with Ellen Williams and her group\textcolor{black}{, and instructive comments from H.\ van Beijeren}.

\section*{Appendix: Derivation of Eq.~(\ref{eq:fp3}) and Some Consequent Results}
In this appendix we expand the derivation of the Fokker-Planck equation given in Ref.~\onlinecite{pge}, as well as correcting some algebra oversights in intermediate steps presented there.  We also present some simpler expressions for quantities of interest that arise for the case of non-interacting [energetically] steps ($\varrho=2$) investigated in the reported computations.
\setcounter{equation}{0}
\renewcommand{\theequation}{A\arabic{equation}}

As in Ref.~\onlinecite{pge}, we begin with the correspondence found by Dyson between RMT and his Coulomb gas model \cite{D62}:
$N$ classical particles on a line, interacting with a logarithmic potential, and confined by an overall harmonic potential. Dyson's model helps our
understanding of the fluctuation properties of the spectrum of complex conserved systems. This model can be generalized to the \textit{dynamic} Brownian motion model, in which the $N$ particles are subject, besides the mutual Coulomb repulsions, to dissipative
forces \cite{guhr}. The particle positions $x_i$ then obey Langevin equations,
\begin{equation}
\dot x_i = -\gamma x_i + \sum_{i\neq j}\frac{\hat{\varrho}}{x_i-x_j} +\sqrt{\Gamma} \eta,
\label{eq:Langevin}
\end{equation}
where $\eta$ is a delta-correlated white noise and $\hat{\varrho}$ ($\propto \varrho$) is the ``charge" of each particle. The probability of finding the particles at the positions $\{x_n\}$ at time $t$ is the solution of the multidimensional FPE
\begin{eqnarray}
\label{eq:fpex}
{\partial P(\{x_n\},t)\over\partial t}&=&\sum_i{\partial \over\partial x_i}\left[{\partial \over\partial
x_i}P(\{x_n\},t)+ \gamma x_i P(\{x_n\},t)\right] \nonumber \\ &-& \sum_{i\neq j}{\partial \over\partial x_i}\left[\frac{\hat{\varrho}}{x_i-x_j}
P(\{x_n\},t)\right].
\end{eqnarray}
In the 1D case, $\gamma^{-1}$ would essentially be the variance of the stationary distribution. Narayan and Shastry \cite{NS93} showed that the CS model is
equivalent to Dyson's  Brownian motion model, in the sense that the solution of the FPE (\ref{eq:fpex}) may be written as
$P(\{x_n\},t) \! = \! \psi(\{x_n\},t)\psi_0(\{x_n\},t)$, where $\psi(\{x_n\},t)$ is the solution of a Schr\"odinger equation with imaginary time, derived from the CS Hamiltonian.   The \textcolor{black}{deterministic} force of Eq.~(\ref{eq:Langevin})

\begin{equation}
{F}(x_m)=- \gamma x_m -\sum_{k>m}{\hat{\varrho}\over x_k-x_m}+\sum_{q<m}{\hat{\varrho}\over x_m-x_q},
\end{equation}
so that

\begin{eqnarray}
\hspace{-2mm} {F}(x_{m \! + \! 1}) \! - \! {F}(x_m) = - \! \gamma (x_{m \! + \! 1} \! -
 \!  x_m)  - \hat{\varrho}
\left[{- 2 \over x_{m \! + \! 1} \! - \! x_m}  \right.
\label{eq:Fx}
\end{eqnarray}
\begin{eqnarray}
\left. \! +\! \sum_{k>m \! + \! 1}\! {x_{m+1}- x_m\over (x_k \! - \! x_{m \! + \! 1})(x_k \! - \! x_m)}+\! \sum_{q<m}{x_{m+1}- x_m\over
(x_{m \! + \! 1} \! - \! x_q)(x_m \! - \! x_q)}\right]. \nonumber
\end{eqnarray}

Our goal is to find the distribution of widths $w$.  Mindful of the Gruber-Mullins 
approach \cite{GM}, we construct a single-``particle," mean-field approximation in which the dynamical variable is the nearest-neighbor distance
$w_m\equiv x_{m+1}-x_m$. To decouple the force on $w_m$ from the other particles, we assume---in the
spirit of GM---that the denominators $(x_k-x_{m+1})(x_k-x_m)$ in Eq.~(\ref{eq:Fx}) are replaced by their mean values, the average being taken in the stationary state:
\begin{equation}
\langle (x_k-x_{m\! + \! 1})(x_k\! - \! x_m) \rangle_{st}=\langle w^2\rangle_{st}(k\! - \! m\! - \! 1)(k\! - \! m),
\label{eq:mean}
\end{equation}

\noindent Each of the two sums in Eq.~(\ref{eq:Fx}) then simplifies greatly, taking the form \begin{equation}
{(x_{m+1}- x_m) \over \langle w^2\rangle_{st}}\times \left(\sum_{p=1}^N {1 \over (p+1)p} = {N \over N+1}   \; \raisebox{-1.5ex}{$\stackrel{\textstyle \rightarrow}{\scriptstyle N \rightarrow \infty}$} \; 1\right).
\end{equation}
Hence, the interaction of a particle pair with all other
particles acts on average as a harmonic potential, increasing the ``spring constant" of the external confining potential. We arrive at a single-particle Langevin
equation for the terrace width $w$:
\begin{equation}
\label{eq:L1}
\frac{dw}{dt}=-2\left[\left(\frac{\gamma}{2} +\frac{\hat{\varrho}}{\langle w^2\rangle_{st}}\right) w -{\hat{\varrho}\over w}\right]+\sqrt{2\Gamma}\eta.
\end{equation}

Our goal is to convert Eq.~(\ref{eq:L1}) into a FPE for which Eq.~(\ref{eq:Ps}) is a steady-state solution. We change to dimensionless variables $s \equiv w/\langle w\rangle_{st}$ and $\tilde t \equiv \Gamma t/\langle w\rangle_{st}^2$.  Treating $\gamma$ as a self-consistency parameter and recognizing $\hat{\varrho}= \varrho \Gamma/2$, we set $\gamma =  \Gamma / \langle w^2\rangle_{st}$.  Then the coefficient in parentheses in Eq.~(\ref{eq:L1}) becomes

\begin{equation}
\frac{(1+\varrho)\Gamma}{2\langle w^2\rangle_{st}} = \frac{b_\varrho \Gamma}{\langle w\rangle_{st}^2}.
\end{equation}

\noindent using the second moment of $P_\varrho(s)$ [$\langle s^2\rangle\! =\! (\varrho+1)/(2b_\varrho)$].  Furthermore, if $\langle \eta(t) \eta(t^\prime)\rangle = \delta(t-t^\prime)$, then $\tilde\eta(\tilde{t}) \equiv \sqrt{2/\Gamma}\langle w\rangle_{st} \eta(t)$ satisfies $\langle \tilde\eta(\tilde{t}) \tilde\eta(\tilde{t}^\prime)\rangle = \delta(\tilde{t}-\tilde{t}^\prime)$. With these results, we recast Eq.~(\ref{eq:L1}) into the Langevin equation

\begin{equation}
\label{eq:Ls}
\frac{ds}{d\tilde t}=-\left[2b_\varrho s -\frac{\varrho}{s}\right]+ \tilde\eta.
\end{equation}

\noindent and thence the sought-after FPE given in Eq.~(\ref{eq:fp3}).

To solve Eq.~(\ref{eq:fp3}) we must specify the initial distribution in $s_0$. For an initial (at $\tilde{t}=0$) sharp distribution $\delta (s\! - \! 1)$, corresponding to a perfectly cleaved crystal, the solution is essentially written down by Montroll and West:\cite{MW,Strat}

\begin{equation}
\hspace{-0.5mm} P(s,\tilde t)\! =\! 2\tilde{b}_\varrho \, s^{\frac{\varrho\! +\!1}{2}} \,
{\rm e}^{\frac{(\varrho\! -\!1) \tilde t}{4}}
I_{\frac{\varrho\! -\!1}{2}}\left(  2\tilde{b}_\varrho s {\rm e}^{-\frac{\tilde t}{2}}\right){\rm e}^{-\tilde{b}_\varrho (s^2 \! +\! {\rm e}^{-\tilde t})},
\label{eq:Pdelt}
\end{equation}
\noindent where $\tilde{b}_\varrho \!  \equiv \! b_\varrho/(1\! -\! {\rm e}^{-\tilde t})$.  In the limit of long times, we showed in Ref.~\onlinecite{pge} that, as $\tilde t$ increases, this $P(s,\tilde t)$ approaches $P_\varrho(s)$ of Eq.~(\ref{eq:Ps}).

For the particular case $\varrho$=2, $\tilde{b}_\varrho$ becomes $4/[\pi(1\! -\! {\rm e}^{-\tilde t})]$, while $I_{\frac{1}{2}}(z) = \sqrt{2/(\pi z)} \, \sinh(z)$.  Then Eq.~(\ref{eq:Pdelt}) simplifies to

\begin{eqnarray}
\hspace{-0.5mm} P(s,\tilde t)\! = \! \frac{2^{3\over 2} \,
{\rm e}^{\frac{3 \tilde t}{4}}\, s}{\pi \sinh^{1\over 2}(\frac{\tilde t}{2})}
\sinh\!\left(\! \frac{(4/\pi) s}{\sinh(\frac{\tilde t}{2})}\!\right)\!\exp\! \left[\!  -\frac{4 (s^2 \! +\! {\rm e}^{-\tilde t})}{\pi(1\! -\! {\rm e}^{-\tilde t})}\! \right].
\label{eq:Pdelt2}
\end{eqnarray}

In experiments, $P(s)$ is generally characterized just by its variance $\sigma^2 \! \equiv \! \mu_2 \! -\! \mu_1^2$, which can be calculated from its second and first moments, $\mu_2$ and $\mu_1$, respectively:

\begin{equation}
\mu_2(\tilde{t}) = \frac{3\pi}{8}u_{\tilde{t}} + {\rm e}^{-\tilde{t}}
= \left(\frac{3\pi}{8}-1\right)\left(1-{\rm e}^{-\tilde{t}}\right) + 1
\label{eq:mu2}
\end{equation}

\begin{equation}
\hspace{-2mm}\mu_1(\tilde{t})  \! = \! \frac{1}{2}\left[
u_{\tilde{t}}^{\frac{1}{2}}\exp \! \left(\frac{-4/\pi}{{\rm e}^{\tilde{t}}-1}\right)  \!  +  \!  \left\{\! 1\! + \! \left(\! \frac{8}{\pi}\! -\! 1 \! \right){\rm e}^{-\tilde{t}}\! \right\} \!
\Upsilon(\tilde{t})\right] \! \! \!
\label{eq:mu1}
\end{equation}

\noindent where, for brevity, we take $u_{\tilde{t}} \equiv 1 \! - \! \exp(-\tilde{t})$, which obviously approaches unity exponentially from below. 
Furthermore, we write $\Upsilon(\tilde{t})\! \equiv \! (\pi/4)\exp(\tilde{t}/2)\, {\rm erf}\! \left(2/[\pi (\exp(\tilde{t})-1)]^{1/2}\right)$, where  
erf is the error function;\cite{AbSt} $\Upsilon(\tilde{t})$ also approaches unity exponentially, but from above, after rising initially from $\pi/4$ to about 1.01.

Scrutiny of Eq.~(\ref{eq:mu1}) reveals that each of the two summands in the square brackets approaches 1 for large $\tilde{t}$.  As $\tilde{t}$ approaches 0, the first summand vanishes while the second rises to 2.  Thus, $\mu_1(\tilde{t})$ has the expected value for vanishing and large $\tilde{t}$, as illustrated in the inset of Fig.~\ref{f:tau-mu1}.  There is, however, an initial rapid drop, reaching a minimum of about 0.9745 around $\tilde{t}$=0.582, and then rising smoothly, reaching 0.99 by $\tilde{t}$=1.95, 0.995 by $\tilde{t}$=2.685, and 0.999 by $\tilde{t}$=4.33. The small deviation from unity is \textcolor{black}{presumably} due to the approximations in using Eq.~(\ref{eq:mean}) \textcolor{black}{to reach Eq.~(\ref{eq:L1}), which apparently break the symmetry of the fluctuations of the steps ($m$ and $m\! +\! 1$) bounding $w_m$ \cite{Meannote}.}

To the extent that this deviation is negligible  [and in any case for qualitative purposes], we get 
\begin{equation}
\sigma^2(\tilde{t})|_{\mu_1 \! \equiv \! 1} \! = \! \sigma^2_W (1 - {\rm e}^{-\tilde{t}}).
\label{eq:sigcleave}
\end{equation}
\noindent If we numerically evaluate $\sigma^2(\tilde{t})$ using Eqs.~(\ref{eq:mu2}) and (\ref{eq:mu1}), we find a similar expression but with a more rapid rise to the equilibrium result; remarkably, it is well approximated by
\begin{equation}
\sigma^2(\tilde{t}) \! = \! \sigma^2_W (1 - {\rm e}^{-2\tilde{t}}).
\label{eq:sigmu}
\end{equation}

\noindent
reminiscent of the solution of the Fokker-Planck equation for a Brownian particle in a quadratic potential \cite{MW2}.  In any case, the Arrhenius behavior of the characteristic time of the exponential will not be affected by such modest changes in the prefactor.

If the Fokker-Planck description step relaxation is robust, then the higher moments of $P(s,\tilde{t})$ should also characterize those moments extracted from the KMC data, as displayed in the text in Fig.~\ref{f:mu2340}.  Thus, we present explicit analytic formulas for the third and fourth moments:
\newcounter{aequation}
\renewcommand{\theequation}{A\arabic{equation}\alph{aequation}}
\begin{eqnarray}
\label{eq:mu3}
\mu_3(\tilde{t}) &=& \left(\frac{1}{2} {\rm e}^{-\tilde{t}}u_{\tilde{t}}^{1/2}+\frac{5\pi}{16} u_{\tilde{t}}^{3/2}\right)\exp \! \left(-\frac{4/\pi}{{\rm e}^{\tilde{t}}-1}\right) \nonumber \\ &&+\left(\frac{4}{\pi} {\rm e}^{-2\tilde{t}} 
+ 3 u_{\tilde{t}}{\rm e}^{-\tilde{t}} + \frac{3 \pi}{16} u_{\tilde{t}}^2\right)\Upsilon(\tilde{t}) \\
 \addtocounter{equation}{-1}\addtocounter{aequation}{1}
&\approx& 1 + \left(\frac{\pi}{2}-1\right)\left(1-{\rm e}^{-\tilde{t}/\textsc{t}_3}\right)
\label{eq:mu3a}
\end{eqnarray}
\addtocounter{aequation}{-1}
\begin{eqnarray}
\label{eq:mu4}
\mu_4(\tilde{t}) &=& \frac{15 \pi^2}{64} u_{\tilde{t}}^2 + \frac{5 \pi}{4}{\rm e}^{-\tilde{t}}u_{\tilde{t}} +  {\rm e}^{-2\tilde{t}} \\ \addtocounter{equation}{-1} \addtocounter{aequation}{1}
&\approx& 1 + \left(\frac{15}{64}\pi^2-1\right)\left(1-{\rm e}^{-\tilde{t}/\textsc{t}_4}\right) 
\label{eq:mu4a}
\end{eqnarray}

The approximate expressions for $\mu_3(\tilde{t})$ and $\mu_4(\tilde{t})$ in Eqns.~(\ref{eq:mu3a}) and (\ref{eq:mu4a}) are written in the form of the exact result for $\mu_2(\tilde{t})$ in Eq.~(\ref{eq:mu2}).  By setting $\textsc{t}_3 = \textsc{t}_4 = 1$ one obtains a mediocre approximation which underestimates $\mu_3(\tilde{t})$ and $\mu_4(\tilde{t})$ by as much as 6\% and 15\%, respectively.  A far better accounting is obtained by taking $\textsc{t}_3 \approx 0.79$ and $\textsc{t}_4 \approx 0.76$; the best values of these time constants depends weakly on the temporal range over which one seeks to optimize the agreement.  The approximate expressions then underestimate the actual $\mu_j(\tilde{t})$ (by at most 2\% and 3\%) up to $\tilde{t} \approx 3/2$ and then overestimate it (by at most $\frac{1}{2}$\% and 1\%), respectively.  Thus, $\mu_3(\tilde{t})$ and $\mu_4(\tilde{t})$ can be well described by curves starting rising smoothly from unity and decaying exponentially toward their long-time limit, but with values of $\textsc{t}_j$ that are smaller than unity.  Finally, note that the approximate expressions based on Eqns.~(\ref{eq:mu3a}) and (\ref{eq:mu4a}) are not used in the analysis of the moments in Subsection \ref{s:HiMo}; hence, the values of the $\textsc{t}_j$ play no role there.

From these results the skewness can be expressed analytically but has an unwieldy form.  However, it is semiquantitatively described by 0.4857 tanh($\tilde{t}$) (i.e.\ to within $\pm 4\%$ for $\tilde{t} \ge 0.46$ and within a percent for $\tilde{t} \ge 1.9$).  In other words, the skewness rises smoothly and monotonically from 0 initially to the equilibrium value.  Since for large $\tilde{t}$, $\tanh (\tilde{t}) \sim 1 - 2 \exp (-2\tilde{t})$ we find the same approach to saturation as for the variance in Eq.~(\ref{eq:sigmu}).
The kurtosis begins at 3 but dips (to about 2.87 near $\tilde{t}$=0.5) before rising to its equilibrium value of 3.1082. The approach to this asymptotic value is well approximated by $3.11\, (1 \! -\! 0.58 {\rm e}^{-2\tilde{t}})$.


\begin{thebibliography}{99}
\bibitem{time} The mean of the Poisson-distributed update intervals [\onlinecite{Voter}] was about 0.01 sec.  We saved when the counter reached 100 times this mean.  The mean overshoot was also about 0.01 sec.
\bibitem{nu4} The factor of 4 accounts for the 4 directions in which an atom can move from a general position on the surface.  In some treatments it is absorbed into $\nu_0$.
\bibitem{polarmis} For initial states small polar misorientations and so with small densities of kinks with one orientation, equilibration of the step shape still involves
creation of new kinks, and this in turn implies detachment of 3-bonded atoms, since detachment of an atom from a kink does not change the number of kinks.  For larger misorientations and kink densities, more subtle effects---with profound implications---come into play.  Since these findings distract from the focus of this Communication, we defer discussion to a separate paper. [\onlinecite{kiss}]
\bibitem{MW2} Specifically, Eq.~(3.21) of Ref.~\onlinecite{MW} shows that the variance of the Gaussian distribution is proportional to $1 \! - \! \exp(-2t)$.
\end{thebibliography}
\end{document}